%% file: paper.tex
\documentclass[%
 reprint,
 superscriptaddress,
 nofootinbib, 
 nobibnotes,
 amsmath,amssymb,
 aps,
pdftex,
]{revtex4-1}

\usepackage[dvipdfmx]{graphicx}
\usepackage{dcolumn}
\usepackage[hypertexnames=false, backref=page]{hyperref}
\usepackage{enumerate}
\usepackage{bm}
\usepackage{comment}
\usepackage{url}
\usepackage{siunitx}
\usepackage{color}
\usepackage{tikz}
\usepackage{mathtools}
\usepackage{docmute}

\usepackage[utf8x]{inputenc}
\usepackage[T1]{fontenc}

\newcommand{\eq}[2][]{
	\begin{align#1}
		#2
	\end{align#1}
}

\newcommand{\diff}{\mathrm{d}}
\newcommand{\iu}{\mathrm{i}}
\newcommand{\e}{\mathrm{e}}
\newcommand{\sinc}{{\rm{sinc}}}

\newcommand{\figref}[1]{Fig.~\ref{#1}}
\newcommand{\tabref}[1]{Table~\ref{#1}}
\newcommand{\secref}[1]{Sec.~\ref{#1}}
\newcommand{\appref}[1]{Supplemental material}

\newcommand{\av}[1]{\left\langle #1 \right\rangle}

\newcommand{\unit}[1]{\, {\rm #1}}

\allowdisplaybreaks[4] 

\definecolor{mygreen}{RGB}{0,130,0} 

\begin{document}


\title{Observational constraint on axion dark matter with gravitational waves}

\author{Takuya Tsutsui}
\affiliation{%
Research Center for the Early Universe (RESCEU), Graduate School of Science, The University of Tokyo, Tokyo 113-0033, Japan
}
\affiliation{%
Department of Physics, Graduate School of Science, The University of Tokyo, Tokyo 113-0033, Japan
}
\author{Atsushi Nishizawa}%
\affiliation{%
Research Center for the Early Universe (RESCEU), Graduate School of Science, The University of Tokyo, Tokyo 113-0033, Japan
}
\email{anishi@resceu.s.u-tokyo.ac.jp}

\date{\today}

\begin{abstract}
Most matter in the Universe is invisible and unknown, and is called dark matter.
A candidate of dark matter is the axion, which is an ultralight particle motivated as a solution for the $CP$ problem.
Axions form clouds in a galactic halo, and amplify and delay a part of gravitational waves propagating in the clouds.
The Milky Way is surrounded by the dark matter halo composed of a number of axion patches.
Thus, the characteristic secondary gravitational waves are always expected right after the reported gravitational-wave signals from compact binary mergers.
In this paper, we derive a realistic amplitude of the secondary gravitational waves.
Then we search the gravitational waves having the characteristic time delay and duration with a method optimized for them.
We find no significant signal.
Assuming the axions are a dominant component of dark matter, we obtain the constraints on the axion coupling to the parity-violating sector of gravity for the mass range, [$1.7 \times 10^{-13}, 8.5 \times 10^{-12}$]$\,\unit{eV}$, which is  at most $\sim 10$ times stronger than that from Gravity Probe B.

\end{abstract}

\maketitle



\section{Introduction} \label{SEC:introduction}
Most matter in the Universe is invisible and is called dark matter.
Many candidates have been considered for dark matter.
One candidate is a quantum chromodynamics (QCD) axion~\cite{peccei_quinn} which is a pseudo Nambu-Goldstone boson introduced to resolve the $CP$ problem.
The $CP$ problem is that many experiments, especially the measurements of the neutron electric-dipole moment~\cite{electric_dipole_moment}, prefer that the electric charge conjugate $\mathcal{C}$ and the parity $\mathcal{P}$ symmetries are conserved, which requires fine tuning in QCD theory.
Furthermore, the existence of axions is also expected from the string theory~\cite{string_axiverse}, and is called the string axion.
The mass of the string axion ranges widely because it depends on the way of the compactification of the extra dimensions occurs.
Thus, we should search for a broad mass range for both axions.

Although axions were historically introduced in the standard model of particle physics, axions have often been searched for with electromagnetic (EM) interaction in laboratory experiments~\cite{Irastorza:2018dyq,Sikivie:2020zpn}, and in astrophysics through the observations of supernovae~\cite{axion_EM_supernovae} and active galactic nuclei~\cite{axion_EM_AGN}. Also the cosmological evolution of axions has been studied~\cite{cosmological_axion1, cosmological_axion2, cosmological_axion3, cosmological_axion4, cosmological_axion5, cosmological_axion6, cosmological_axion7, cosmological_axion8, cosmological_axion9}. Many search methods and recent constraints are reviewed in~\cite{review_axion_EM, DiLuzio:2020wdo, Galanti:2022ijh}.

Axions can also interact with gravity.
Then, similar to the coupling to an electromagnetic field tensor in the QCD axion Lagrangian, we consider the Chern-Simons (CS) interaction term coupled to axions; that is, the axion nonminimal coupling to a Riemann tensor~\cite{review_CSgravity}.
The additional term is the simplest coupling of a pseudoscalar field to gravity (the one appears in the CS gravity) which is a low-energy effective theory of the parity-violating extension of GR~\cite{review_CSgravity}.
The CS coupling of axions have been searched for in~\cite{axion_CMB, axion_large_scale_structure, axion_PTA1, axion_PTA2, axion_PTA3, axion_IFO1, axion_IFO2, axion_binary, axion_BH1, axion_BH2, axion_nuclear_spin_precession, soda-urakawa, Ali-Haimoud:2011zme}.
In the previous study~\cite{Ali-Haimoud:2011zme}, the measurement of the frame-dragging effect by Gravity Probe B around the Earth constrained the coupling, $\ell \lesssim 10^8 \unit{km}$.

If axions interact with gravity through the CS term, gravitational waves (GWs) induce axion decay into gravitons~\cite{soda-yoshida}.
If dark matter in the Milky Way (MW) is composed of axions, the propagating GWs are amplified and delayed.
Since axions are expected to be cold dark matter in the MW halo, the GWs from the axion decay are almost monochromatic.
Therefore, axions through the interaction generate characteristic secondary GWs whose features depend on the axion mass and the coupling to the parity-violating sector of gravity~\cite{soda-urakawa}.

In this paper, we search the monochromatic secondary GWs induced by primary GWs from coalescences of binary neutron stars (BNSs) and binary black holes (BBHs) in the observational data of GW detectors.\footnote{Recently the possibility to search for axion dark matter through neutrino oscillation and a stochastic GW background was pointed out in~\cite{axion_constraint_NFW}.}
Our method is optimized for the time delay and the signal duration.
Then, from no detection of the axion signal, we constrain the coupling constant, improving the upper limit from Gravity Probe B.

The organization of this paper is as follows.
In \secref{SEC:property}, we review the properties of the secondary GWs generated by axion decay.
We show the method we analyze GW data in \secref{SEC:method} and give the results in \secref{SEC:result}.
We discuss the results and future prospects in \secref{SEC:discussion}.
Finally, \secref{SEC:conclusion} is devoted to a summary.

In this paper, we use the natural units $\hbar=c=1$.

\section{Properties of secondary gravitational waves} \label{SEC:property}
In this section, we briefly review the axion decay and enhancement of a GW in the MW halo, following Jung et al.~\cite{soda-urakawa}, and enumerate the characteristic quantities for axions that we use later in the search.
The Lagrangians are
\eq{
	S =& S_{\mathrm{EH}} + S_{a} + S_{\mathrm{CS}}, \label{EQ:total_action} \\
	S_{\mathrm{EH}} =& \frac{1}{16 \pi G} \int d^4 x \sqrt{-g} R, \\
	S_a =& - \int d^4 x \sqrt{-g} \left( \frac{1}{2} \nabla_\mu a \nabla^\mu a + \frac{m^2_a}{2} a^2 \right), \\
	S_{\mathrm{CS}} =& \frac{\ell^2}{16 \sqrt{2 \pi G}} \int d^4 x \sqrt{-g}  a {^\ast\!} R R,
}
where $\ell$ is the axion coupling constant in the parity-violating sector of gravity, $m_a$ is the axion mass, $a$ is the axion field, and $^*RR$ is the Pontryagin density.

\subsection{Characteristic quantities}
A GW propagating through the MW halo composed of the axion dark matter with mass $m_a$ produces secondary GWs which are almost monochromatic.
The resonance frequency $f_\mathrm{res}$ is
\eq{
	f_\mathrm{res} &= \frac{1}{2\pi} \frac{m_a}{2} = 12 \unit{Hz} \left( \frac{m_a}{10^{-13} \unit{eV}} \right) \;, \label{EQ:fpeak}
}
and its width is
\eq{
\varDelta f_\mathrm{res} =& 2 f_\mathrm{res} \varDelta v \nonumber \\ 
	=& 2.4 \times 10^{-2} \unit{Hz} \left( \frac{m_a}{10^{-13} \unit{eV}} \right) \left( \frac{\varDelta v}{10^{-3}} \right) \;,
}
where $\varDelta v$ is a velocity dispersion of dark matter~\cite{MW_dark_matter_velocity_dispersion}.
From Eq.~\eqref{EQ:fpeak}, the LIGO~\cite{LIGO1, LIGO2} sensitive band, \SI{20}{Hz} -- \SI{1024}{Hz}, corresponds to $\SI{1.7e-13}{eV} < m_a < \SI{8.5e-12}{eV}$. 
Therefore around a detected GW from compact binary coalescence (CBC), we can find the GW spectrum with a peak at only $f_\mathrm{res}$. The signal duration of the secondary GWs is roughly given by the inverse of the line width,
\eq{
	\varDelta t_\mathrm{duration} =& \frac{1}{\varDelta f_\mathrm{res}} = \frac{2\pi}{m_a \varDelta v} \nonumber \\
		=& \SI{41}{s} \left( \frac{10^{-13} \,\unit{eV}}{m_a} \right) \left( \frac{10^{-3}}{\varDelta v} \right)\;.
}

The MW is composed of many coherent clouds of axion dark matter whose size is about the coherent length $L_\mathrm{c}$ defined by
\eq{
	L_{\rm c} \coloneqq& \frac{2\pi}{m_a\varDelta v} \nonumber \\ 
		=& \SI{4.0e-7}{pc} \left( \frac{10^{-13} \unit{eV}}{m_a} \right) \left( \frac{10^{-3}}{\varDelta v} \right) \;. \label{EQ:coherent_length}
}
We note that the factor of $2\pi$ is different from the definition in~\cite{soda-urakawa}.
Then, the following quantities including $L_c$ are modified from those in~\cite{soda-urakawa} by the factor of $2\pi$.

The time-averaged group velocity of a GW in the axion clouds, $\bar{v}_g$, is
\eq{
	\bar{v}_g = 1 - \frac{1}{3} \left( \frac{\sqrt{\pi G \rho_a} \ell^2 m_a^2}{2} L_\mathrm{c} \right)^2 \label{EQ:group_velocity}.
}
Assuming that the MW halo is spherical symmetric and its radius is $R \sim \SI{100}{kpc}$, we can derive the delay time of the secondary GWs from the primary GW,
\eq{
	\varDelta t_\mathrm{delay} =& \frac{R}{\bar{v}_g} - \frac{R}{c} \nonumber \\
		\simeq & \frac{\pi^3 R}{3} \frac{ G \rho_a \ell^4 m_a^2}{\varDelta v^2} \nonumber \\
		=& \SI{1.1e3}{s} \left( \frac{R}{100 \unit{kpc}} \right) \left( \frac{10^{-3}}{\varDelta v} \right)^2 \left( \frac{\ell_\mathrm{eff}}{10^8 \unit{km}} \right)^4 \nonumber\\
		&\times \left( \frac{m_a}{10^{-13} \unit{eV}} \right)^2 \left( \frac{\rho_\mathrm{DM}}{0.3 \unit{GeV/cm^3}} \right). \label{EQ:Delta_t}
}
where $\rho_a$ and $\rho_\mathrm{DM}$ are the energy densities of axions and dark matter.
We defined the effective coupling parameter,
\eq{
	\ell_\mathrm{eff} \coloneqq \ell f_\mathrm{DM}^{1/4} \label{EQ:l_eff} \;,
}
including the fraction of the amount of axions to the total amount of dark matter, $f_\mathrm{DM} \coloneqq \rho_a / \rho_\mathrm{DM}$.

\begin{figure}[tb]
	\includegraphics[width=0.9\linewidth]{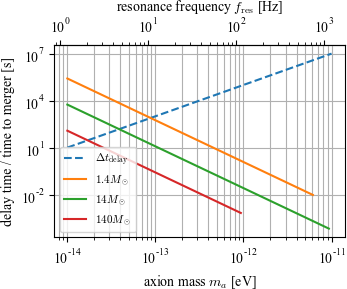}
	\caption{The signal delay $\varDelta t_\mathrm{delay}$ (dashed line) and the time to merger for various equal-mass binaries (solid lines) as a function of $m_a$ for $\ell_\mathrm{eff} = \SI{e8}{km}$, $R = \SI{100}{kpc}$, $\varDelta v = 10^{-3}$, and $\rho_\mathrm{DM} = \SI{0.3}{GeV/cm^3}$. Solid lines are plotted up to the ISCO frequency of a Schwarzchild BH for BNS of \SI{1.4}{M_\odot}--\SI{1.4}{M_\odot} and their merger frequencies for other BBH.}
	\label{FIG:delay_duration}
\end{figure}

Figure~\ref{FIG:delay_duration} is an example for the signal delay and duration compared with the time to merger for various compact binaries.
When the delay is longer than the time to merger, the axion signal is observed after the binary merger ($m_a \gtrsim 10^{-13}\unit{eV}$ in \figref{FIG:delay_duration}).

\subsection{Amplitude enhancement of secondary GWs} \label{SEC:amplification_model}
We are surrounded by the MW dark matter halo with size $R$.
Since axions form clouds~\cite{soda-urakawa} with the $L_c$, the MW halo contains many axion patches.
The number of patches, $N$, along the propagation path of a GW is estimated to be
\eq{
N =& R / L_c \nonumber \\
=& 2.5 \times 10^{11} \left( \frac{R}{100\unit{kpc}} \right) \left( \frac{\varDelta v}{10^{-3}} \right) \left( \frac{m_a}{10^{-13}\unit{eV}} \right) \;.
}

In the MW halo, GWs from CBC induce the decay of axions into monochromatic GWs at the frequency, $f_\mathrm{res}$.
In other words, the propagating GW is amplified in each axion patch by a factor of $1 + \delta_\mathrm{patch}$~\cite{soda-urakawa}, where
\eq{
	\delta_\mathrm{patch} = \frac{1}{2} \left( \frac{\sqrt{\pi G\rho_a} \ell^2 m_a^2}{2} L_\mathrm{c} \right)^2 \sinc^2 \left( 2\pi \frac{f - f_\mathrm{res}}{\varDelta f_\mathrm{res}} \right) . \label{EQ:delta_patch}
}
The authors of~\cite{soda-urakawa} neglect the phase rotation of the GW propagating through each axion patch because it is tiny.
In this case, the amplification in each patch is treated coherently and the GW amplitude grows exponentially, giving the amplification factor of $[(1 + \delta_\mathrm{patch}) \e^{\iu \theta}]^N \simeq \e^{N\delta_\mathrm{patch}}$ for sufficiently large $N$ and tiny phase shift $\theta$ for the secondary GW (see~\appref{SEC:review_pahse_shift}).
However, the statement is not true for the entire MW halo; because the number of patches $N$ is very large and the tiny phase shift $\theta$ is accumulated during propagation in the MW halo, the total phase shift finally reaches the limit beyond which the GW is no longer enhanced coherently.
Such imperfect amplification has been numerically simulated in the time domain in~\cite{fujita-yamada}, but for the first time we derive the analytical formula below.

Since a secondary GW is almost monochromatic with the width $\varDelta f_\mathrm{res}$, let us consider the two frequency modes at the edges of the resonance, that is, at $f_\mathrm{res} + \varDelta f_\mathrm{res}$ and $f_\mathrm{res} - \varDelta f_\mathrm{res}$.
If the phase difference accumulates over $\pi$, those modes suppress each other incoherently.
Then, the phase shift is accumulated coherently during the period $\varDelta t_{\rm coh}$ defined by $2\pi (2\varDelta f_\mathrm{res}) \varDelta t_{\rm coh} = \pi$; that is, $\varDelta t_{\rm coh} = 1/(4 \varDelta f_\mathrm{res}) = L_{\rm c}/4$.
When propagating in $\varDelta t_\mathrm{coh}$, a secondary GW passes the number of patches $N_\mathrm{c}$:
\eq{
	N_{\rm c} \coloneqq N \frac{\varDelta t_{{\rm coh}}}{\varDelta t_\mathrm{delay}} \;,
}
because the delay from one axion patch is $\varDelta t_\mathrm{delay} / N$.
If $N_{\rm c} > N$, the enhancement occurs coherently in all patches. However, in the case of $N_{\rm c} < N$, the resonant growth stops after passing $N_{\rm c}$ patches and the total enhancement is given by an incoherent superposition of $N / N_{\rm c}$ enhanced GWs.
The total enhancement factor is given by
\eq{
    F_{\rm halo} \simeq \left\{
        \begin{array}{ll}
            \displaystyle 
            \e^{N \delta_\mathrm{patch}} \qquad \qquad {\rm for}\; N_{\rm c} \geq N \\ 
            \displaystyle
            \e^{N_{\rm c} \delta_\mathrm{patch}} \sqrt{\frac{N}{N_{\rm c}}} \quad \; {\rm for}\; N_{\rm c} < N 
        \end{array} \right. \;. \label{eq:F-total}
}
with
\eq{
    N \delta_\mathrm{patch} =& \frac{\pi^2 G}{4} \frac{Rm_a^3 \ell^4 \rho_a}{\varDelta v} \av{\sinc^2\left(2\pi \frac{f - f_\mathrm{res}}{\varDelta f_\mathrm{res}} \right)} \\
        =& 9.3 \left( \frac{10^{-3}}{\varDelta v} \right) \left( \frac{R}{100\unit{kpc}} \right) \left( \frac{m_a}{10^{-13}\unit{eV}} \right)^3 \nonumber \\
        & \left( \frac{\ell_\mathrm{eff}}{10^8\unit{km}} \right)^4 \left( \frac{\rho_\mathrm{DM}}{0.3\unit{GeV}/\unit{cm}^3} \right) \;, \label{eq:Ndel} \\
    N_{\rm c} \delta_\mathrm{patch} =& \frac{3}{8} \;. \label{eq:Ncdel}
}
where the $\sinc$ factor is averaged for $|f - f_\mathrm{res}| < \varDelta f_\mathrm{res}$ because GWs are amplified in the narrow range, $\av{\sinc^2\left( 2\pi \frac{f - f_\mathrm{res}}{\varDelta f_\mathrm{res}} \right)} = 0.24$.
The critical case is $N=N_{\rm c}$, i.e., from Eqs.~(\ref{eq:Ndel}) and (\ref{eq:Ncdel}), the critical coupling is
\eq{
    \ell_c^\mathrm{eff} =& 4.5 \times 10^7 \unit{km} \left( \frac{\varDelta v}{10^{-3}} \right)^{1/4} \left( \frac{100\unit{kpc}}{R} \right)^{1/4} \nonumber \\
    & \left( \frac{10^{-13}\unit{eV}}{m_a} \right)^{3/4} \left( \frac{0.3\unit{GeV}/\unit{cm}^3}{\rho_\mathrm{DM}} \right)^{1/4} \label{EQ:critical_coupling} \;.
}

\begin{figure*}[tb]
    \begin{center}
    \raisebox{0.8mm}{\includegraphics[width=7.6cm]{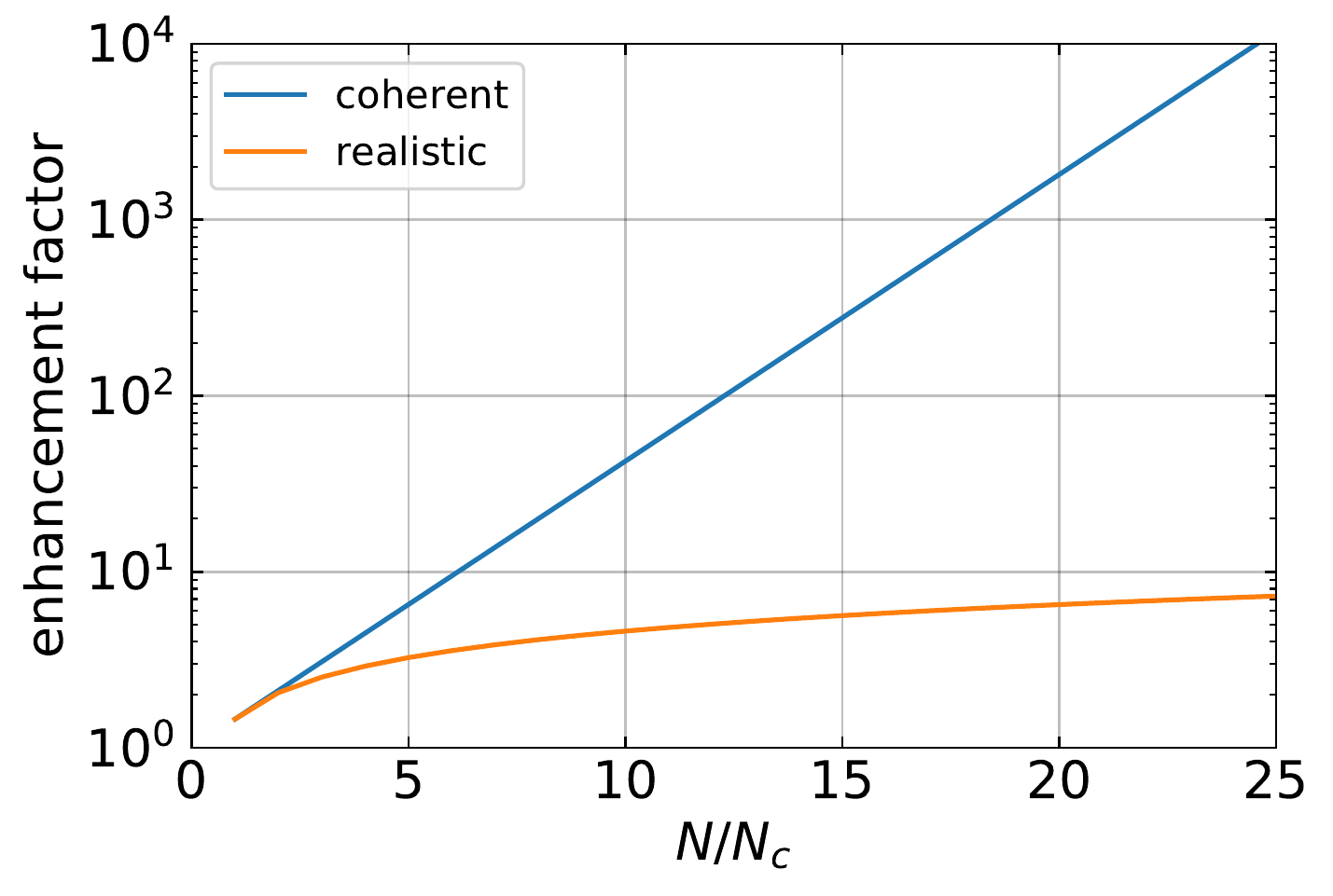}}
    \includegraphics[width=7.3cm]{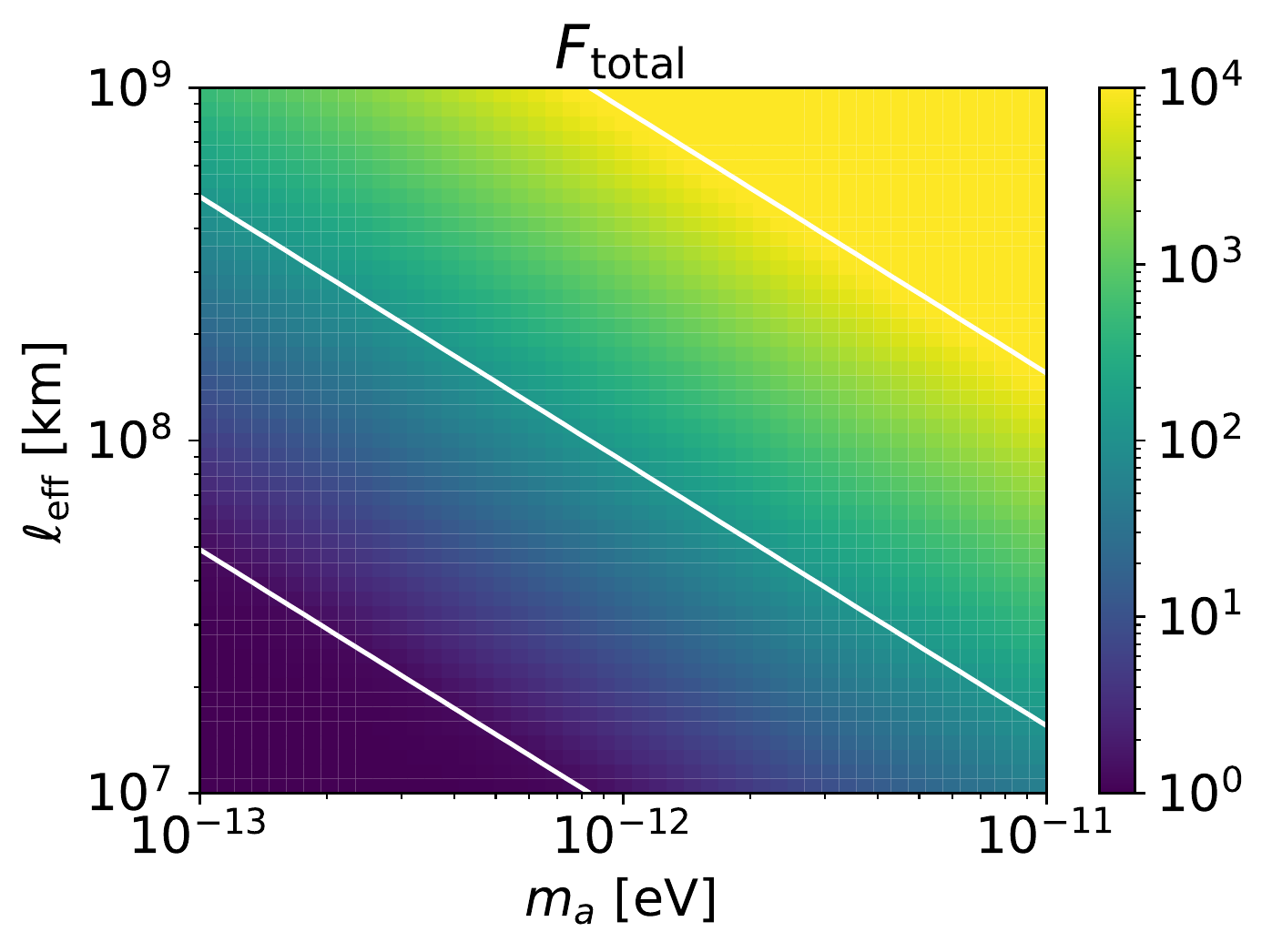}
    \caption{Left: the coherent ($\theta=0$) (blue) and realistic (orange) enhancement factors as a function of $N/N_{\rm c}$. Right: the realistic enhancement factor as a function of $m_a$ and $\ell_\mathrm{eff}$. The white diagonal lines from the bottom to the top are the coupling strengths of $N/N_{\rm c}=1$ (the critical coupling), $10^4$, and $10^8$.}
    \label{fig:incoherent-enhancement}
    \end{center}
\end{figure*}

In the left panel of Fig.~\ref{fig:incoherent-enhancement}, the coherent ($\theta=0$) and realistic enhancement factors are plotted as a function of $N/N_{\rm c}$ for illustration. It is shown that neglecting the phase shift $\theta$ in each axion patch significantly overestimates the enhancement factor for $N>N_{\rm c}$. In the right panel of Fig.~\ref{fig:incoherent-enhancement}, the realistic enhancement factor in Eq.~(\ref{eq:F-total}) is plotted as a function of $m_a$ and $\ell_\mathrm{eff}$. Even for the realistic case, the  enhancement factor can be significantly large for large $N/N_{\rm c}$ (the upper-right region).

\section{Method} \label{SEC:method}
In this section, a search method optimized for axion signals is explained.
If an axion signal is in data, it should be almost a monochromatic wave at the frequency of $f_\mathrm{res}$ for a time duration $\varDelta t_\mathrm{duration}$ starting at $\varDelta t_\mathrm{delay}-\tau(f_\mathrm{res})$ from the coalescence time, where the time to merger is $\tau(f) = (5/256) (\pi f)^{-8/3} \left( G \mathcal{M}/c^3 \right)^{-5/3}$ and $\mathcal{M}$ is the chirp mass.
For the axion mass range that we search, $\SI{1.7e-13}{eV} < m_a < \SI{8.5e-12}{eV}$, the time to merger is negligible compared with the delay time so that the starting time of the secondary GW is almost at $\varDelta t_\mathrm{delay}$ from the coalescence time.
The Fourier amplitude of the secondary signal is given by
\eq{
	\left| \tilde{h}_\mathrm{axion}(f_\mathrm{res}) \right| = \left[ F_\mathrm{halo}(f_\mathrm{res}) - 1 \right] \left| \tilde{h}_\mathrm{CBC}(f_\mathrm{res}) \right| \;, \label{EQ:relation_CBC_and_axion_amplitude}
}
where $\tilde{h}_\mathrm{CBC}(f)$ is the Fourier amplitude of a primary GW signal from a CBC.
For $\tilde{h}_\mathrm{CBC}$ of a BBH, we use the IMRPhenomD waveform~\cite{PNexample_3.5PN_1, PNexample_3.5PN_2}, which is an aligned spinning inspiral-merger-ringdown waveform, setting the high frequency cutoff to the peak frequency at which the amplitude of the waveform is maximized. While for BNS, the waveform is not accurate enough for high frequencies because of tidal deformation and the high-frequency cutoff is set to the innermost stable circular orbit (ISCO) frequency for a Schwarzchild black hole (BH).
When obtaining the CBC amplitude, we need the distance to the GW source and the antenna responses at the time of each event.
For conservative constraints, the farthest distance within the $1\sigma$ error is used~\cite{GWOSC}. As discussed in~\appref{SEC:antenna_responses}, the difference of the antenna responses hardly affects the results.
The events used in this paper are enumerated later, and the data are downloaded from~\cite{GWOSC}.
The other waveform parameters (e.g. the chirp mass, inclination angle, right ascension, declination, and so on) are set to the maximum-likelihood values.

To search the axion signals, we use the following steps:
\begin{enumerate}
	\item Make the $\chi^2_\mathrm{obs}$ map on the $m_a$--$\ell_\mathrm{eff}$ plane from whitened data:\\ $\chi^2_\mathrm{obs} \coloneqq \sum_{I \in \mathrm{IFO}} \left| \tilde{d}_I(f_\mathrm{res}) \sqrt{\varDelta f} / \sqrt{S_{n,I}(f_\mathrm{res})} \right|^2$,
	where $\tilde{d}_I$ is the Fourier amplitude of the data starting at $\varDelta t_\mathrm{delay}$ from the coalescence time with the chunk size of $\varDelta t_\mathrm{duration}$, $S_{{\rm n},I}(f)$ is the power spectral density (PSD) for the $I$ th detector, and $\varDelta f$ is the width of a frequency bin.
	\begin{enumerate}
		\item Take a chunk of whitened data with the duration equal to $\varDelta t_\mathrm{duration}(m_a)$ from $\varDelta t_\mathrm{delay}(m_a, \ell_\mathrm{eff})$ for a set of the parameters $(m_a, \ell_\mathrm{eff})$. \label{ITEM:chunk}
		\item Calculate the $\chi^2_\mathrm{obs}$ at $f_\mathrm{res}$. \label{ITEM:chi2}
		\item Repeat (\ref{ITEM:chunk}) and (\ref{ITEM:chi2}) for other $(m_a, \ell_\mathrm{eff})$.
	\end{enumerate}
	\item Make a $p$-value map on the $m_a$--$\ell_\mathrm{eff}$ plane from the $\chi^2_\mathrm{obs}$ map.
		The $p$-value to reject the $\ell_\mathrm{eff}$ is
		\eq{
			p(m_a, \ell_\mathrm{eff}) \coloneqq \int_0^{\chi^2_\mathrm{obs}} p_\chi \big(x \big| \overline{\chi^2} \big) \,\diff x
		}
		where $\overline{\chi^2} \coloneqq \sum_{I\in\mathrm{IFO}} |\tilde{h}_\mathrm{axion}(f_\mathrm{res}) \sqrt{\varDelta f} / \sqrt{S_{n,I}(f_\mathrm{res})}|^2$.
		For example, the $p$-value is tiny when the expectation value of the noncentral $\chi^2$-distribution is much larger than the actually observed one.
	\item Search for parameter sets, $(m_a, \ell_\mathrm{eff})$, for which the $p$-value is larger than a threshold, that is, judge if the axion signal exists.
		We set the threshold to $5\times 10^{-3}$ or at $0.5\%$ credible level.
	\item Combine the search results from multiple GW events to check consistency among all GW events.
		Logical OR of nondetection is used to combine the results.
		That is, if the parameter sets are rejected even by one GW event, they are also rejected in the combined result.
		Then we derive upper limits on $\ell_\mathrm{eff}$ for each $m_a$.
\end{enumerate}

As the constraints are combined for multiple GW events, the constraint on the $\ell_\mathrm{eff}$ becomes always tighter with more events.
However, if the axion signal exists, the true parameter set cannot be rejected even if the number of the combined GW events is large enough.

\begin{table*}[tb]
	\centering
	\caption{Properties of primary GWs analyzed in this paper. The frequency cutoff is set to \SI{1024}{Hz} for the peak frequency $> \SI{1024}{Hz}$.}
	\begin{tabular}{lcccccc} \hline\hline
		\shortstack{Event \\name}& \shortstack{Primary \\mass [$M_\odot$]} & \shortstack{Secondary \\mass [$M_\odot$]} & \shortstack{Frequency \\cutoff [Hz]} & \shortstack{Network SNR of \\a primary GW} & \shortstack{Duty cycle of \\ the data used} & \shortstack{References} \\ \hline
		GW170814               & 31  & 25  & \SI{5.9e2}{} & 16 & 92\% & \cite{GWTC-2, GW170814} \\
		GW170817               & 1.5 & 1.3 & \SI{8.0e2}{} & 33 & 79\% & \cite{GWTC-2, GW170817_observation, GW170817_multimessenger} \\
		GW190728\_064510 & 12  & 8.1 & \SI{1.0e3}{} & 14 & 61\% & \cite{GWTC-3} \\
		GW200202\_154313 & 10  & 7.3 & \SI{1.0e3}{} & 11 & 97\% & \cite{GWTC-3} \\
		GW200316\_215756 & 13  & 7.8 & \SI{1.0e3}{} & 10 & 100\% & \cite{GWTC-3} \\ \hline
	\end{tabular}
	\label{TAB:event_list}
\end{table*}

Since the GW detectors, LIGO and Virgo~\cite{Virgo}, are sensitive at \SI{20}{Hz}--\SI{1024}{Hz}, we search the corresponding mass range, $\SI{1.7e-13}{eV} < m_a < \SI{8.5e-12}{eV}$,
in this paper.
However, the axion signal does not exist beyond the peak frequency because no primary GW exists as in Eq.~\eqref{EQ:relation_CBC_and_axion_amplitude}.
Thus, for each GW event, we search the axion signal up to the peak frequency.
However, since we use the IMRPhenomD waveform up to the ISCO frequency for BNS, the cutoff frequency GW170817 is given by the ISCO frequency.

For this search to constrain the wide parameter range, we need the data that last longer after the amplitude peak of primary GWs and that are available from all detectors (LIGO-Hanford, LIGO-Livingston, and Virgo).
We require the length of the data to be about \SI{5.3e4}{s} for GW170814~\cite{GW170814}, GW170817~\cite{GW170817_observation, GW170817_multimessenger}, and GW190728\_064510~\cite{GWTC-2}, and about \SI{1.2e4}{s} for GW200202\_154313 and GW200316\_215756~\cite{GWTC-3}, which are determined from computational time and the lower fraction of lacking data (see \appref{SEC:tricks}).
Furthermore, for the third observing run (O3) events, we require the network signal-to-noise ratio (SNR) to be higher than 10 and the peak frequency higher than \SI{1024}{Hz}.
The GW events satisfying the conditions are listed in~\tabref{TAB:event_list}.

\section{Results} \label{SEC:result}
\begin{figure}[tb]
    \centering
    \includegraphics[width=0.8\linewidth]{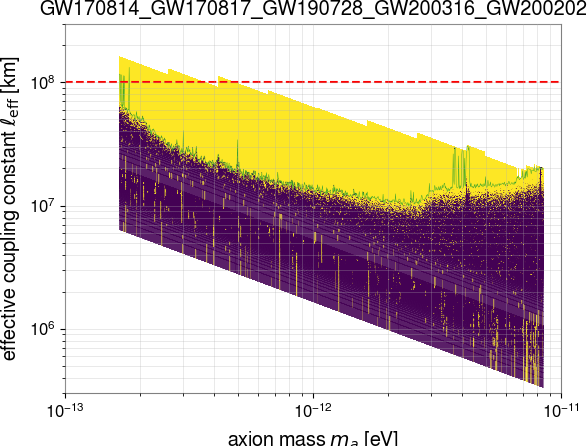}
    \caption{Constraint on the $\ell_\mathrm{eff}$ as a function of $m_a$.  The yellow region is rejected at more than $99.5\%$ level, the purple one is not excluded, and the white one is not searched. The green line is an upper limit and the red dashed one is the previous constraint from Gravity Probe B for $f_\mathrm{DM} = 1$.}
    \label{FIG:result_GW170814_GW170817_GW190817_GW200202_GW200316}
\end{figure}

\begin{figure}[tb]
    \centering
    \includegraphics[width=0.75\linewidth]{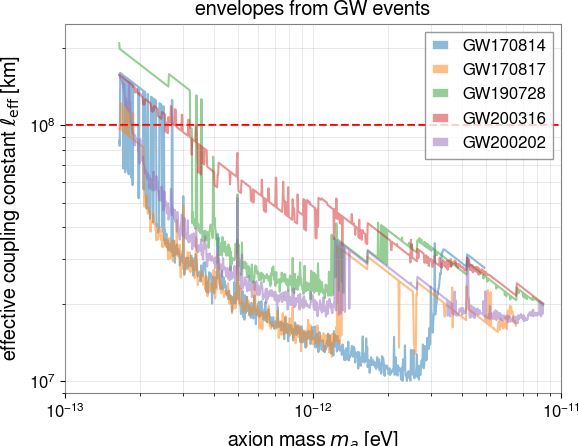}
    \caption{Minimum values of $\ell_{\rm eff}$ constrained at more than $99.5\%$ level from each GW event in~\tabref{TAB:event_list}. The red dashed line is the previous constraint from Gravity Probe B for $f_\mathrm{DM} = 1$.}
    \label{FIG:result_each_GW170814_GW170817_GW190817_GW200202_GW200316}
\end{figure}

Figure~\ref{FIG:result_GW170814_GW170817_GW190817_GW200202_GW200316} shows the search result from the events listed in ~\tabref{TAB:event_list}.
The yellow region is excluded at the $99.5\%$ credible level, but the purple is not.
The white region is not searched---the reason is discussed in the next section.
The green line is an upper limit of the constraint, which is the maximum of $\ell_\mathrm{eff}$ not excluded in the coupling parameter searched.
Since there are many points sparsely distributed on the $(m_a, \ell_\mathrm{eff})$ plane that were not ruled out due to accidental noise fluctuations, we consider the envelope to obtain conservative upper limits on the coupling $\ell_\mathrm{eff}$ for each $m_a$.
The upper limit has some peaks; those are caused by detector line noises.
The red dashed line is the previous constraint from Gravity Probe B~\cite{Ali-Haimoud:2011zme}.
The constraint is improved by at most one order of magnitude from the Gravity Probe B.
The upper shape of the yellow region is like a saw.
This is because the search is performed from weaker to stronger couplings until the unsearched $\ell_\mathrm{eff}$ or $\varDelta t_\mathrm{delay}$ is encountered. For larger $\ell_\mathrm{eff}$ or longer time delay, the data chunk searched is far from the event time of the primary GW, and the detector data are more likely to be in the nonscience mode.

Figure~\ref{FIG:result_each_GW170814_GW170817_GW190817_GW200202_GW200316} shows the minimum values of $\ell_\mathrm{eff}$ that were ruled out from each GW event in~\tabref{TAB:event_list}.
In \figref{FIG:result_each_GW170814_GW170817_GW190817_GW200202_GW200316}, there are many step-functionlike jumps of the lines.
This is due to short duration noises. Since $\varDelta t_\mathrm{delay} \propto m_a^2 \ell_\mathrm{eff}^4$, the search with the mass heavier than that at which the jump exists always encounters the short duration noise and gives almost the same largest $\ell_\mathrm{eff}$ not excluded.
On the other hand, in the search with lighter masses, we can neglect the short duration noise because the contribution is diluted enough due to longer duration of the chunks.
Although the highest SNR event is GW170817, the strongest constraint is obtained from GW170814.
This is because, even though the SNR is lower, the amplitude of a primary GW is larger for BBH events, producing a larger secondary GW.

\section{Discussions} \label{SEC:discussion}
The constraints are obtained for the MW dark matter halo parameters, $R = \SI{100}{kpc}, \varDelta v = 10^{-3}$, and $\rho_\mathrm{DM} = \SI{0.3}{GeV/cm^3}$, but these measurement values should have large uncertainties.
Nevertheless, the effects on our results are expected to be small, because the dependence of $\ell_{\rm eff}$ on the halo parameters is the power of $1/4$ from Eq.~\eqref{EQ:critical_coupling}. 
Given the uncertainties of $\mathcal{O}(10\%)$~\cite{review_axion} in the MW dark matter halo parameters, they modify $\ell_\mathrm{eff}$ by the order of $\mathcal{O}(1\%)$.
Although we assume that the halo is homogeneous in this paper, we could consider a realistic halo density profile, which is more dense at the center and coarse at the edge~\cite{NFW_profile, axion_constraint_NFW}.
However, it is beyond the scope of this paper.

Our constraint is stronger than that in~\cite{soda-urakawa}.
Their method is simply to set the condition, $\chi_\mathrm{obs} > 100$, for detection.
The threshold corresponds to $F_\mathrm{total} \sim 100$ in~\figref{fig:incoherent-enhancement}. Our detection threshold is roughly $F_\mathrm{total}\sim 5$--$50$ and much smaller thanks to statistically evaluating the $\chi^2$-value and combining multiple events.  
Thus the difference of the constraints is from that of the detection criteria.

Also, there is another constraint from the observation of a neutron star with NICER~\cite{ILoveQ_nonDynamicalCS_observation}.
Although the constraint looks much stronger than ours, it is not obvious for some assumptions in~\cite{ILoveQ_nonDynamicalCS_observation} to be satisfied for $\ell \sim \SI{e8}{km}$.
Differences between our and their constraints are discussed in~\appref{SEC:NICER}.

Future GW detectors are more sensitive than or have the sensitive frequency bands different from that of the current GW detectors. However, as discussed in~\appref{SEC:future_detectors}, we cannot expect a significant improvement of the constraint on the coupling due to weak dependence on GW amplitude.

\section{Conclusions} \label{SEC:conclusion}
GWs from CBCs are delayed and amplified during the propagation in an axion dark matter halo.
In this paper, we have derived a realistic enhancement of the secondary GW amplitude, taking into account the accumulation of a phase delay during propagation and searched such signals with characteristic duration and time delay in the GW observational data. Since we know the signal duration and the time delay of axion signal, we take the data chunk whose length is the same as the duration at the time delay after a binary merger.
In the search, we use the data right after the five reported GWs from CBCs.
Then, since the signal duration and the time delay depend on the axion mass and the CS coupling, we analyze the data chunk and constraint the coupling by comparing the search result with the expected amplitude of the secondary GW.
The constraint on the effective coupling constant $\ell_\mathrm{eff}$ for axion mass in the range of [$1.65 \times 10^{-13}, 8.47 \times 10^{-12}$]\,$\unit{eV}$ is at most $\sim 10$ times improved from the previous study, Gravity Probe B~\cite{Ali-Haimoud:2011zme}.

\acknowledgments
We thank S.~Morisaki for fruitful discussions and valuable comments on the draft of the paper.
T.~T. is supported by JSPS KAKENHI Grant No. 21J12046. A.~N. is supported by JSPS KAKENHI Grants No. JP19H01894 and No. JP20H04726 and by Research Grants from the Inamori Foundation.
The authors are grateful for computational resources provided by the LIGO Laboratory and supported by National Science Foundation Grants No. PHY-0757058 and No. PHY-0823459.
This material is based upon work supported by NSF's LIGO Laboratory which is a major facility fully funded by the National Science Foundation.



\input{supplement}


\bibliographystyle{unsrt}
\bibliography{references}

\end{document}

%% file: supplement.tex

\title{Supplemental Material: Observational constraint on axion dark matter with gravitational waves}

\author{Takuya Tsutsui}
\affiliation{%
Research Center for the Early Universe (RESCEU), Graduate School of Science, The University of Tokyo, Tokyo 113-0033, Japan
}
\affiliation{%
Department of Physics, Graduate School of Science, The University of Tokyo, Tokyo 113-0033, Japan
}
\author{Atsushi Nishizawa}%
\affiliation{%
Research Center for the Early Universe (RESCEU), Graduate School of Science, The University of Tokyo, Tokyo 113-0033, Japan
}
\email{anishi@resceu.s.u-tokyo.ac.jp}

\date{\today}

\maketitle



\begin{widetext}
\appendix

\section{Brief review of the phase shift of a secondary GW} \label{SEC:review_pahse_shift}
By axion decaying, secondary GWs are generated and superposed the observational GWs has large amplitude and a tiny phase shift~\cite{soda-urakawa}.
However, it is useful to explain more about the tiny phase shift, and then we review it briefly in this section.
You can refer~\cite{soda-urakawa} for details.

If a GW propagate along to $z$-coordinate, it can be written as
\eq{
    h_{ij}(z, t) = h_{ij}^R(z, t) + h_{ij}^L(z, t)
}
where the both helicity modes are expressed as
\eq{
    h_{ij}^{R/L}(z, t) = e_{ij}^{R/L} h_F^{R/L}(t) \e^{\iu \left( kz - \frac{m_a}{2} t \right)} - \iu e_{ij}^{L/R} h_B^{R/L}(t) \e^{\iu \left( -kz - \frac{m_a}{2} t \right)} + \mathrm{c.c.} \;,
}
where $h_F$ and $h_B$ are complex amplitudes for the forward and the backward waves, $e_{ij}^{R/L}$ is the polarization tensor, and $k$ is the wave number.
By solving the equation of motion for $h_{ij}$ from Eq.~\eqref{EQ:total_action}, we obtain a solution after propagating in one axion patch:
\eq{
    h_F^{R/L}(t = L_c) =& h_F^{R/L}(0) (1 + \delta_\mathrm{patch}) e^{-\iu \psi(t = L_c)} \\
    \psi(t) \simeq& \frac{m_a}{2} \epsilon t \left\{ 1 + \frac{1}{2} [\sinc(m_a \epsilon t) - 1] \left( \frac{\gamma}{\epsilon} \right)^2 \right\} \\
    \epsilon =&\frac{k - m_a / 2}{m_a / 2} \\
    \gamma =& \sqrt{\pi G \rho_a} \ell^2 m_a \\
        =& \SI{5.7e-9}{} \left( \frac{\ell}{\SI{e8}{km}} \right)^2 \left( \frac{m_a}{\SI{e-13}{eV}} \right) \left( \frac{\rho_a}{\SI{0.3}{GeV/cm^3}} \right)^{1/2}
}
and $\delta_\mathrm{patch}$ is Eq.~\eqref{EQ:delta_patch}, $L_c$ is Eq.~\eqref{EQ:coherent_length}, $h_B^{R/L}(t) \sim \gamma h_F^{R/L}(t)$, that is, the backward waves are much smaller than the forward waves, and then we neglect the backward waves~\cite{soda-urakawa}.
In the case, the GW waveform is
\eq{
    h_{ij}^{R/L}(z, t = L_c) =& e_{ij}^{R/L} (1 + \delta_\mathrm{patch}) \e^{\iu \theta} \e^{\iu k(z - t)} + \mathrm{c.c.} \;, \\
    \theta =& -\frac{m_a}{2} \epsilon L_c \frac{1}{2} [\sinc(m_a \epsilon L_c) - 1] \left( \frac{\gamma}{\epsilon} \right)^2 \;.
}
Then, the total amplification factor from $N$ patches is
\eq{
    F_\mathrm{halo} = \left[ (1 + \delta_\mathrm{patch}) \e^{\iu \theta} \right]^N
}
Because of $\gamma \ll 1$, the phase shift $\theta$ is tiny, so that the total amplification factor is $(1 + \delta_\mathrm{patch})^N$ in~\cite{soda-urakawa}.
However, in our paper, we consider the effect of the tiny phase shift.

\section{Antenna responses} \label{SEC:antenna_responses}
In this search, we do not consider the antenna response effects, that is, we assume that all GW detectors have the same antenna responses.
For the unconstrained parameters or the data chunks, the antenna responses vary in $\mathcal{O}(10\%)$, and then the errors in the $\ell_\mathrm{eff}$ are in $\mathcal{O}(1\%)$.
Thus our constraint is practically consistent with more realistic analysis without the assumption of the constant response.

\section{Tricks for an efficient search} \label{SEC:tricks}
In this paper, the constraints for $(m_a, \ell)$ are obtained with the method in \secref{SEC:method}.
However, there is a problem how to divide the parameter space $(m_a, l)$.
For a non-optimal division, the calculations are super heavy and cannot be done in reasonable run time.
We should consider efficient way to take bins of the parameters.

\subsubsection{For effective coupling constant}
If bins for $\ell_\mathrm{eff}$ are sampled (log-)uniformly, the chunks to calculate $\chi_\mathrm{obs}^2$ exist on data densely for the lower $\ell_\mathrm{eff}$ but coarsely for the higher $\ell_\mathrm{eff}$ because of $\varDelta t_\mathrm{delay} \propto \ell_\mathrm{eff}^4$.
For the dense case, those chunks search almost same targets for an axion mass, but for the coarse case, almost different targets, which is inefficient.
Hence, the efficient sampling should be that the intervals between the chunks is constant.
Since the SNR of an axion signal is proportional to an overlap between the axion signal and the chunk, the interval is $0.2 \varDelta t_\mathrm{duration}$ when we approve a $10\%$ SNR loss.
That is, we search data at $\alpha \varDelta t_\mathrm{duration}$ after a binary merger where $\alpha \in \mathbb{R}_{\geq 0}$ is uniformly sampled.
In this paper, we search a range $\alpha \in [0.002, 0.2]$ with less than $0.1\%$ SNR loss and $\alpha \in [0.2, \SI{4e4}{}]$ with less than $10\%$ SNR loss.

To relate the search results with $\ell_\mathrm{eff}$, we solve $\varDelta t_\mathrm{delay} = \alpha \varDelta t_\mathrm{duration}$:
\eq{
	\ell_\mathrm{eff} =& \left( \frac{6 \alpha}{\pi^2 G} \frac{\varDelta v}{R\rho_\mathrm{DM} m_a^3} \right)^{1/4} \\
	  =& \SI{4.4e7}{km} \left( \frac{\SI{100}{kpc}}{R} \right)^{1/4} \left( \frac{\varDelta v}{10^{-3}} \right)^{1/4} \left( \frac{\SI{0.3}{GeV/cm^3}}{\rho_\mathrm{DM}} \right)^{1/4} \left( \frac{10^{-13}\,\unit{eV}}{m_a} \right)^{3/4} \left( \frac{\alpha}{1} \right)^{1/4} . \label{EQ:coupling_from_shift_ratio}
}
We can know from Eq.~\eqref{EQ:coupling_from_shift_ratio} how to divide the range of $\ell_\mathrm{eff}$.
Because the effective coupling constant $\ell_\mathrm{eff}$ is proportional to the shift ratio $\alpha^{1/4}$, the bins on the $m_a$--$\ell_\mathrm{eff}$ plan is highly denser for larger $\ell_\mathrm{eff}$.
Thus, this division is not good to search for larger $\ell_\mathrm{eff}$ because of the computational costs.

\subsubsection{For axion mass}
The bin width of the axion mass $m_a$ is considered in this sub-section.
Since the axion mass is related to the resonance frequency $f_\mathrm{res}$ with Eq.~\eqref{EQ:fpeak}, the mass bin width is interpreted as a frequency bin width.
Since the frequency bin width should be equal to $\varDelta f_\mathrm{res}$, the bin width of the axion mass is log-uniform: $\varDelta\log m_a = \varDelta v$.
However, to obtain the $\chi^2_\mathrm{obs}$, we have to do Fourier transform for the each mass bin, because the length of the data chunk is assigned with $m_a$.
It needs a large calculation cost although the analyzed chunks are almost same between $m_a$ and $m_a + \varDelta m_a$.
This is inefficient, and then we group the some mass bins.
When the axion signal duration is $10\%$ shorter than the chunk size, the SNR is $10\%$ diluted.
That is, approving $10\%$ SNR loss, we can use the same chunk for a group $g_m = \{0.1m + n\varDelta\log m_a | n, m\in\mathbb{Z}, 0 \leq n\varDelta\log m_a < 0.1\}$.
The length of the chunk is the longest one in $g_m$.
By this grouping, we can make the calculation $10\% / \varDelta v = 100 \left( 10^{-3}/\varDelta v \right)$ times rapider.

\section{Computational costs} \label{SEC:computational_costs}
Since the upper limit from the Gravity Probe B~\cite{Ali-Haimoud:2011zme} is $\ell \sim 10^8 \unit{km}$, we should search all the region below $\ell_\mathrm{eff} \lesssim 10^8 \unit{km}$.
However, in~\figref{FIG:result_GW170814_GW170817_GW190817_GW200202_GW200316}, the stronger region of $\ell_\mathrm{eff}$ for $m_a \gtrsim \SI{e-12}{eV}$ is not searched because of computational cost.
Since the delay is larger in the unsearched region because of $\varDelta t_\mathrm{delay} \propto \ell_\mathrm{eff}^4 m_a^2$,
We have to take finer bins for larger $\ell_\mathrm{eff}$ not to lose SNR of an axion signal.
Concretely, to obtain the constraint up to $\ell_\mathrm{eff} = \SI{e8}{km}$ at the heaviest axion mass of $\SI{8.5e-12}{eV}$, we need the computational time of $\sim \SI{20}{years}$, which is $10^3$ times longer than the current calculation (see \appref{SEC:tricks}).
Also, the other problem is that there exists no data at large time delay, because the delay for $\ell_\mathrm{eff} = \SI{e8}{km}$ and $m_a = \SI{8.5e-12}{eV}$ is about \SI{9.3}{months} but the GW observations have already ended at the time.
Thus, searching all the unsearched region below $\ell_\mathrm{eff} = \SI{e8}{km}$ is not feasible with state-of-the-art technology and data.

\section{Constraints from NICER} \label{SEC:NICER}
There is another constraint from the observation of a neutron star (NS) with NICER~\cite{ILoveQ_nonDynamicalCS_observation}.
For massless CS gravity or at the massless limit of the axion-CS coupling ~\cite{ILoveQ_nonDynamicalCS_observation}, the NS observation gives the significantly stronger constraint, $\ell \lesssim \SI{10}{km}$.
The constraint for the massless CS gravity might be applied also to the massive or axion CS case with the mass range as low as the de Broglie wavelength is much longer than the size of a NS. However, the applicability is nontrivial because the constraint is obtained under the small coupling approximation~\cite{ILoveQ_nonDynamicalCS_theory}, which is obviously invalid for the coupling range we searched, $\SI{10}{km} \lesssim \ell \lesssim 10^8 \unit{km}$.
Also, there is another study for scalar-tensor theories~\cite{scalar-tensor_NS-mass-radius}, which states that a constraint for a massive scalar field is much worse than that for a massless one.
This is because the effects of spontaneous scalarization and finite mass are compensated each other in the equation of motion of a scalar field. As a consequence of the degeneracy between the constrained coupling parameter and the mass of the scalar field, the posterior distribution is elongated toward the stronger regime of the coupling parameter.
The scalar-tensor theory is, of course, different from the CS gravity.
However, from the similarity of the equation of motions between the scalar field in~\cite{ILoveQ_nonDynamicalCS_theory} and in~\cite{scalar-tensor_NS-mass-radius}, we expect that the statement of the worse constraint might be true also in the case of the CS gravity.

\section{Future detectors} \label{SEC:future_detectors}
\subsection{ground-based}
Although the future ground-based GW detectors~\cite{ET_paper, ET_science, CE1, CE2} are about ten times sensitive, it improves the constraint of $\ell_\mathrm{eff}$ by a factor of a few because of $\ell_\mathrm{eff} \propto \mathrm{SNR}^{-1/2}$, where SNR is that for the primary GW, from Eqs.~\eqref{eq:F-total} and~\eqref{EQ:relation_CBC_and_axion_amplitude} for an incoherent case, which is the regime of the current upper limit.
On the other hand, as the sensitivities of GW detectors are improved, many new GW events will be detected.
From \figref{FIG:result_each_GW170814_GW170817_GW190817_GW200202_GW200316}, combining the search results from multiple events is obviously important to improve the constraint.

\subsection{space-based}
With space-based GW detectors~\cite{LISA_paper, DECIGO1, DECIGO2} and pulsar timing array~\cite{PTA}, we can search lower frequencies or lighter axion masses.
However, the sensitivities to the $\ell_\mathrm{eff}$ are much worse than those of ground-based ones at lower frequencies.
This is because the scaling, $\ell_\mathrm{eff} \propto m_a^{-3/4} \{ | \tilde{h}_\mathrm{CBC}(f_\mathrm{res}) | \varDelta f_\mathrm{res} \}^{-1/2}$, is obtained from Eqs.~(\ref{eq:F-total}) and \eqref{EQ:relation_CBC_and_axion_amplitude}. Using $\tilde{h}_\mathrm{CBC}(f) \propto f^{-7/6}$ and given the same PSD of a detector, we have the scaling with the axion mass,
\eq{
\ell_\mathrm{eff} &\propto m_a^{-3/4} f_{\rm res}^{1/12} \propto m_a^{-2/3} \;.
}
Therefore, the constraint would be worse than that from Gravity Probe B at low frequencies.

\end{widetext}


\bibliographystyle{unsrt}
\bibliography{references}